\documentclass{article}

\usepackage{arxiv}

\usepackage[utf8]{inputenc} 
\usepackage[T1]{fontenc}    
\usepackage{booktabs}       
\usepackage{amsfonts}       
\usepackage{nicefrac}       
\usepackage{microtype}      
\usepackage{amsmath}
\usepackage{amssymb}
\usepackage{graphicx}

\title{Fast recursive reconnection and the Hall effect: Hall-MHD simulations}

\author{
  Chen Shi \\
  Earth, Planetary, and Space Sciences\\
  University of California, Los Angeles\\
  Los Angeles, CA 90095, USA \\
  \texttt{cshi1993@g.ucla.edu} \\
   \And
  Anna Tenerani \\
  Department of Physics\\
  The University of Texas at Austin\\
  TX 78712, USA \\
  \AND
  Marco Velli \\
  Earth, Planetary, and Space Sciences\\
  University of California, Los Angeles\\
  Los Angeles, CA 90095, USA \\
  \AND
  San Lu \\
  Earth, Planetary, and Space Sciences\\
  University of California, Los Angeles\\
  Los Angeles, CA 90095, USA \\
}

\begin{document}
\maketitle

\begin{abstract}
Magnetohydrodynamic (MHD) theory and simulations have shown that reconnection is triggered via a fast ``ideal" tearing instability in current sheets whose inverse aspect ratio decreases to  $a/L\sim S^{-1/3}$, with $S$ is the Lundquist number defined by the half-length $L$ of the current sheet (of thickness $2a$). Ideal tearing, in 2D sheets, triggers a hierarchical collapse via stretching of X-points and recursive instability. At each step, the local Lundquist number decreases, until the subsequent sheet thickness either approaches kinetic scales or the Lundquist number becomes sufficiently small. Here we carry out a series of Hall-MHD simulations to show how the Hall effect modifies recursive reconnection once the ion inertial scale is approached. We show that as the ion inertial length becomes of the order of the inner, singular layer thickness at some step of the recursive collapse, reconnection transits from the plasmoid-dominant regime to an intermediate plasmoid+Hall regime and then to the Hall-dominant regime. The structure around the X-point, the reconnection rate, the dissipation property and the power spectra are also modified significantly by the Hall effect.
\end{abstract}


\section{Introduction}\label{sec:intro}
Magnetic reconnection is one of the most important processes in astrophysical plasmas. By allowing changes in the topology of the magnetic field it allows conversion of magnetic energy into the thermal and kinetic energies of the ambient plasma as well as particle acceleration. The study of magnetic reconnection started in the 1950s, when \cite{sweet1958} and \cite{parker1957} established the resistive, stationary, MHD model known as the Sweet-Parker current sheet model, where a uniform resistivity is responsible for reconnecting magnetic field lines carried into a current sheet by a plasma inflow. The thickness (or the aspect ratio) of the current sheet is determined by the resistivity, that by locally diffusing the inwardly convected magnetic field, allows a stationary state to be achieved. However, because astrophysical plasmas are typically nearly-collsionless, Lundquist numbers are enormous, and Sweet-Parker type current sheets are very thin, limiting the speed of  reconnection to extremely small values: defining the Lundquist number of the Sweet-Parker current sheet as $S=LV_A/\eta$ where $2L$, $V_{A}$ and $\eta$ are the length of the current sheet, the asymptotic upstream Alfv\'en speed and the magnetic diffusivity respectively, the inverse aspect ratio $a/L$ and the normalized reconnection rate $R=u_{in}/V_{A}$ scale as $a/L\sim S^{-1/2}$ and $R \sim S^{-1/2}$, where $2a$ is the thickness of the current sheet and $u_{in}$ is the inflow speed. For nearly-collisionless plasma, $S \rightarrow \infty $ and as a result $a/L, R \rightarrow 0$. Thus the Sweet-Parker model is unable to explain the explosive phenomena observed in space and astrophysical contexts, such as solar and stellare flares and coronal mass ejections (CMEs). A more promising resistive MHD model for fast reconnection was proposed by \cite{petschek1964} who hypothesized that a small diffusion region should develop inside the large scale background with an exhaust outflowing from the x-point with a much larger opening angle, leading to a fast reconnection rate $R \sim \left( \ln S \right)^{-1}$. However, resitive-MHD simulations since showed that the Petschek model cannot be achieved with a uniform resistivity, but rather requires a region of enhanced dissipation at the x-point \cite{biskamp1986}.

With the improvements of computational capabilities, numerical studies of magnetic reconnection have reached beyond the MHD regime, exploring Hall-MHD - to remain within the fluid description-, as well as hybrid and direct particle-in-cell (PIC) simulations \cite[e.g.]{birn2001}. The main result stemming from such works is that once two-fluid effects (i.e. the Hall effect) are included, a  fast reconnection rate is reached with a value $R \sim 0.1$, regardless of the specific numerical model used. Many PIC simulations have been carried out \cite[e.g.]{daughton2006} over the last two decades and our understanding of the electron-scale physics has been improved significantly not only thanks to these numerical simulations but also to new observations from the Magnetospheric Multiscale (MMS) mission, whose data provides insight into the electron diffusion region (EDR) near the reconnection site, i.e. the X-point (or X-line) \cite{burch2016}. Interestingly, the increase in resolution of fluid simulations has also lead to new discoveries, altering the resistive-MHD theory reconnection. After \cite{biskamp1986} first noted that at high enough $S$ the Sweet-Parker sheet appeared to become unstable,  \cite{shibata2001} proposed the idea of ``fractal reconnection'': the macroscopic current sheet, due to tearing instability, breaks into multiple X-points, which then dynamically lengthen to form secondary current sheets. These secondary current sheets then become tearing unstable themselves as secondary plasmoids are generated. This recursive process proceeds until some microscopic mechanism stops it. \cite{loureiro2007} showed that the linear growth rate of the tearing mode inside Sweet-Parker type current sheets, i.e. current sheets whose inverse aspect ratio obeys $a/L \sim (LV_A/\eta)^{-1/2}=S^{-1/2}$, scales with a positive power of the Lundquist number: $\gamma \tau_A \sim S^{1/4}$ in the limit $S \rightarrow \infty$ where $\tau_A = L/V_A$ is the Alfv\'en crossing time. \cite{puccivelli2014} carried out more generalized scaling analysis and showed that the current sheet can only thin to $a/L \sim S^{-1/3}$ at which point the growth rate of the tearing mode becomes independent of the Lundquist number (the so called ``ideal tearing'') with growth rates of order 1 ($\gamma \tau_A \sim 1$). Note that in the analysis by \cite{shibata2001}, the same result $a/L \sim S^{-1/3}$ was derived by arguing that the current sheet becomes tearing unstable only when the linear growth time of the fastest-growing mode $t \approx 1/\gamma_{max} $ becomes shorter than the time it takes for the Alfv\'enic outflow to evacuate the reconnected field from the x-point,  $t\sim L/V_A$. \cite{tenerani2015} carried out 2D resistive-MHD simulation of a collapsing macroscopic current sheet and developed the recursive reconnection model further, finding some differences with respect to to \cite{shibata2001}. They also confirmed that the tearing instability happens when a current sheet becomes as thin as $a/L \sim S^{-1/3}$. These results are summarized in \cite{Singetal2019}.

One question within the recursive reconnection picture is what dynamical process intervenes to stop the generation of ever smaller scales: one possible mechanism is the stabilization of the tearing instability due to the decrease of the Lundquist number. It has been shown by early MHD simulations \cite{biskamp1986} that below a threshold $S_c\sim 10^4$ the Sweet-Parker type current sheets are tearing-stable and the inhomogeneous background flow plays the stabilizing role \cite[see]{bulanov1978,shi2018}. Another possible mechanism to stop recursive reconnection is the kinetic effect. In astrophysical and experimental plasmas, e.g. the solar wind, the Earth's magnetotail and the Magnetic Reconnection Experiment (MRX), the ion inertial length $d_i$ can be comparable to the current sheet thickness or the inner layer thickness of the tearing mode \cite[see]{pucci2017}. Even if $d_i$ is much smaller than the thickness of the initial current sheet, the thickness of the current sheets formed during subsequent steps in the recursive collapse may become comparable to $d_i$, well before the Lundquist number reaches the $S_c$ required for stabilization. It has been shown by collisional-PIC and Hall-MHD simulations that when the thickness of a current sheet approaches the ion inertial length, the reconnection rate is enhanced significantly due to the Hall effect \cite{daughton2009,shepherd2010}. Here we carry out a series of Hall-MHD simulations to investigate how recursive reconnection is altered once the Hall term takes effect. We show that, as $d_i$ increases from values smaller to greater than the tearing mode inner, or singular, layer thickness, reconnection transits from a plasmoid-generating dominant pattern to a Hall-dominant pattern. We also show that, during the nonlinear stage, the structure around the X-point, the reconnection rate, the energy dissipation and the power spectra of various quantities are all significantly modified by the Hall effect.

The paper is organized as follows: In Section (\ref{sec:numerical}) we describe the numerical setup of the simulations, including the numerical model and the choice of parameters. In Section (\ref{sec:results}) we show the simulation results and in Section (\ref{sec:summary}) we summarize this work and discuss the possible future works.

\section{Numerical Setup}\label{sec:numerical}
The code we use is a 2.5D  Hall-MHD code based on the following equation set:
\begin{subequations}\label{eq:code_equation}
    \begin{equation}
        \frac{\partial \rho}{\partial t} = - \mathbf{u} \cdot \nabla \rho - ( \nabla \cdot \mathbf{u} ) \rho
    \end{equation}
    \begin{equation}
        \frac{\partial \mathbf{u}}{\partial t} = -  \mathbf{u} \cdot \nabla \mathbf{u} - \frac{1}{\rho} \nabla p + \frac{1}{\rho} \mathbf{j} \times \mathbf{B}
    \end{equation}
    \begin{equation}
        \frac{\partial \psi}{\partial t} = \left( \mathbf{u} \times \mathbf{B} - \frac{d_i}{\rho} \mathbf{j} \times \mathbf{B} \right)_z+ \eta \nabla^2 \psi
    \end{equation}
    \begin{equation}
        \frac{\partial B_z}{\partial t} = \left[ \nabla \times \left( \mathbf{u} \times \mathbf{B} -  \frac{d_i}{\rho} \mathbf{j} \times \mathbf{B} \right) \right]_z + \eta \nabla^2 B_z
    \end{equation}
    \begin{equation}
        \frac{\partial p}{\partial t} = - \mathbf{u} \cdot \nabla p - \kappa (\nabla \cdot \mathbf{u}) p
    \end{equation}
\end{subequations}
where $\mathbf{j} = \nabla \times \mathbf{B}$ is the current density and $\psi$ is the $z$ component (the out-of-plane component) of the magnetic vector potential such that
\begin{equation}
    B_x = \frac{\partial \psi}{\partial y}, \quad B_y = - \frac{\partial \psi}{\partial x}
\end{equation}
and $\nabla \cdot \mathbf{B} = 0$ is conserved automatically. $\kappa = 5/3$ is the adiabatic index. The code is double-periodic and derivatives are calculated by spectral method. Explicit 3rd order Runge-Kutta method is used for time integral and Courant-Friedrichs-Lew condition determines the time step.

The initial condition is a pressure-balanced Harris current sheet:
\begin{subequations}
    \begin{equation}
        \mathbf{B_0} = B_0  \tanh (\frac{y-y_c}{a_0}) \Hat{e}_x
    \end{equation}
    \begin{equation}
        \rho_0 = \rho_\infty + \frac{B_0^2}{2T_0}  \mathrm{sech}^2 (\frac{y-y_c}{a_0})
    \end{equation}
\end{subequations}
where the asymptotic density $\rho_\infty$, the temperature $T_0$ and the asymptotic magnetic field $B_0$ are constants:
\begin{equation}
    \rho_\infty = 1, \quad T_0 =1, \quad B_0 = 1
\end{equation}
$a_0$ is the thickness of the initial current sheet and is set to be 
\begin{equation}
    a_0 = 0.02
\end{equation}
The domain size is $L_x \times L_y = 8 \times 0.8$ and because of the periodicity along $y$, a double Harris current is used so practically $L_y = 0.4$. The number of the uniform grid points is $n_x \times n_y = 4096 \times 2048$, i.e. the resolution is $\Delta x = 1.95 \times 10^{-3}$, $\Delta y = 3.91 \times 10^{-4}$. Random noises of mode $N=1-128$ (i.e. $\mathbf{k_N}=\frac{N}{L_x} \hat{e}_x$) are added to the center of the initial current sheet. Note that ``random'' here means that the phase of each mode is generated randomly but for all the simulations in this work the initial perturbations are exactly the same. The resistivity is $\eta = 1 \times 10^{-5}$, i.e. the Lundquist number $S = 10^{5}$.

7 runs are carried out with the ion inertial length $d_i$  to be the only varying parameter: 
\begin{equation}
    d_i = [0,1,2,3,4,5,10] \times 10^{-3}
\end{equation}
and we will refer to them as Run 0-5 and Run 10 for convenience hereinafter. Note that, the critical thickness of the current sheet to trigger the ``ideal tearing'' is estimated to be
\begin{equation}\label{eq:a_IT}
    a \sim S^{-1/3} = 0.0215
\end{equation}
and the inner layer thickness of the tearing mode is estimated to be 
\begin{equation}\label{eq:innerlayer_IT}
    \delta \sim S^{-1/2} = 3.16 \times 10^{-3}
\end{equation}
Eqs. (\ref{eq:a_IT} \& \ref{eq:innerlayer_IT}) help guide the choice of $a_0$ and $d_i$: $a_0 = 0.02$ allows ideal tearing to be triggered and as $d_i$ varies from values below to above $\delta$  we expect to observe the transition from a plasmoid dominated regime to the Hall regime when an inner layer is formed with a thickness on the order of $d_i$. Run 10 is an extreme case where $d_i$ is on the scale of $a_0$ for comparison.

\section{Results}\label{sec:results}

\subsection{Time evolution and the structure of the current sheet}
\begin{figure}[ht!]
\centering
\includegraphics[scale=0.35]{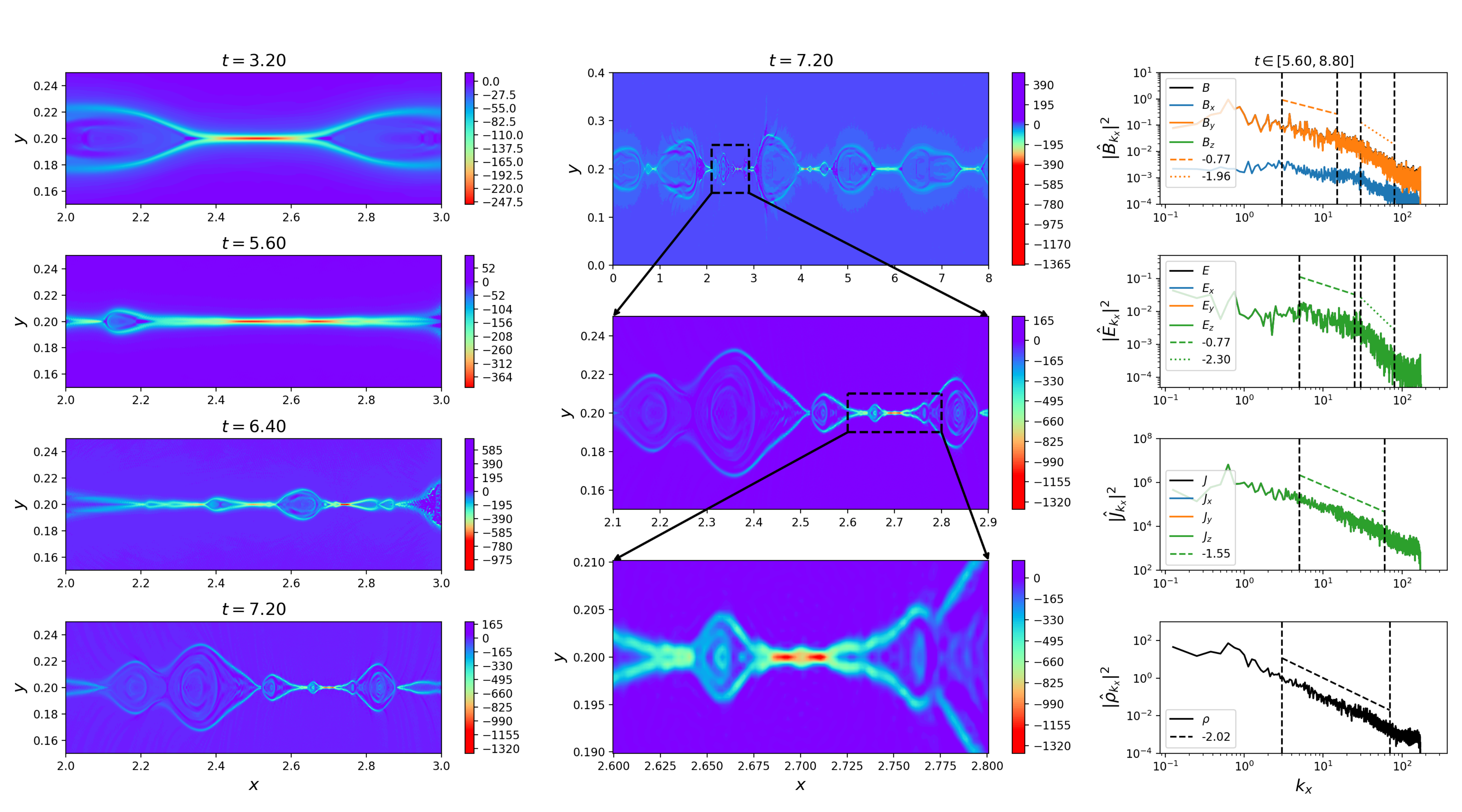}
\caption{Out-of-plane current density $J_z$ and the power spectra in Run 0. Left column: time evolution of $J_z$ inside the domain $x \in[2.0,3.0] $ and $y \in [0.15,0.25]$. Middle column: the hierarchic structure of the current sheet (color is $J_z$) at $t=7.20$. Top panel is the whole simulation domain. Middle and bottom panels are blow-ups as marked by the boxes and arrows. Right column: time-averaged power spectra of the magnetic field $\mathbf{B}$, the electric field $\mathbf{E}$, the current density $\mathbf{J}$ and the density $\rho$ calculated at the central line of the current sheet $y=0.2$ in log-log scale. The blue, orange and green lines represent the $x$, $y$ and $z$ components of the vector fields respectively and the black lines are the sum of the 3 components in the first three panels. The dashed and dotted lines are linear fits for the spectra of $B_y$, $E_z$, $J_z$ and $\rho$. The $k_x$ ranges for the fits are marked by the vertical dashed lines and the slopes of the fits are written in the legends. The time period for averaging the spectra is written at the top of the column. \label{fig:time_series_recur_structure_spectra_0} }
\end{figure}

The left and middle columns of Figure \ref{fig:time_series_recur_structure_spectra_0} show the out-of-plane current density in Run 0, i.e. the MHD case. Left column shows the time evolution of $J_z$ inside the domain $x \in[2.0,3.0] $ and $y \in [0.15,0.25]$ where one X-point appears but note that there are multiple X-points formed along the current sheet. The evolution is similar to what is observed in the simulations by \cite{tenerani2015}. The initial perturbation grows into a sequence of X-points which evolve nonlinearly into secondary current sheets stretched along the outflow direction ($t=3.2$ and $t=5.6$). During the dynamic lengthening, the thickness of the secondary current sheet is observed to be nearly constant and is $a_1 \approx 0.002$, close to the predicted inner layer thickness of the tearing mode. At time $t = 5.6$, secondary tearing is triggered after which the secondary current sheet breaks into a series of small current sheets and plasmoids ($t=6.4$). This process continues recursively until the end of the simulation at $t=9$. The middle column of Figure \ref{fig:time_series_recur_structure_spectra_0} shows the hierarchic structure of $J_z$ at $t=7.20$. Top panel is the whole simulation domain. Middle and bottom panels are blow-ups as marked by the boxes and arrows. It is clearly seen from the bottom panel that at $x \approx 2.70$ a tiny plasmoid is being generated out of a micro current sheet whose thickness is on grid scale: $a_n \approx \Delta y < 0.001$. The right column shows the time-averaged power spectra in this run and will be discussed in Section \ref{sec:power_spectra}.

\begin{figure}[ht!]
\centering
\includegraphics[scale=0.35]{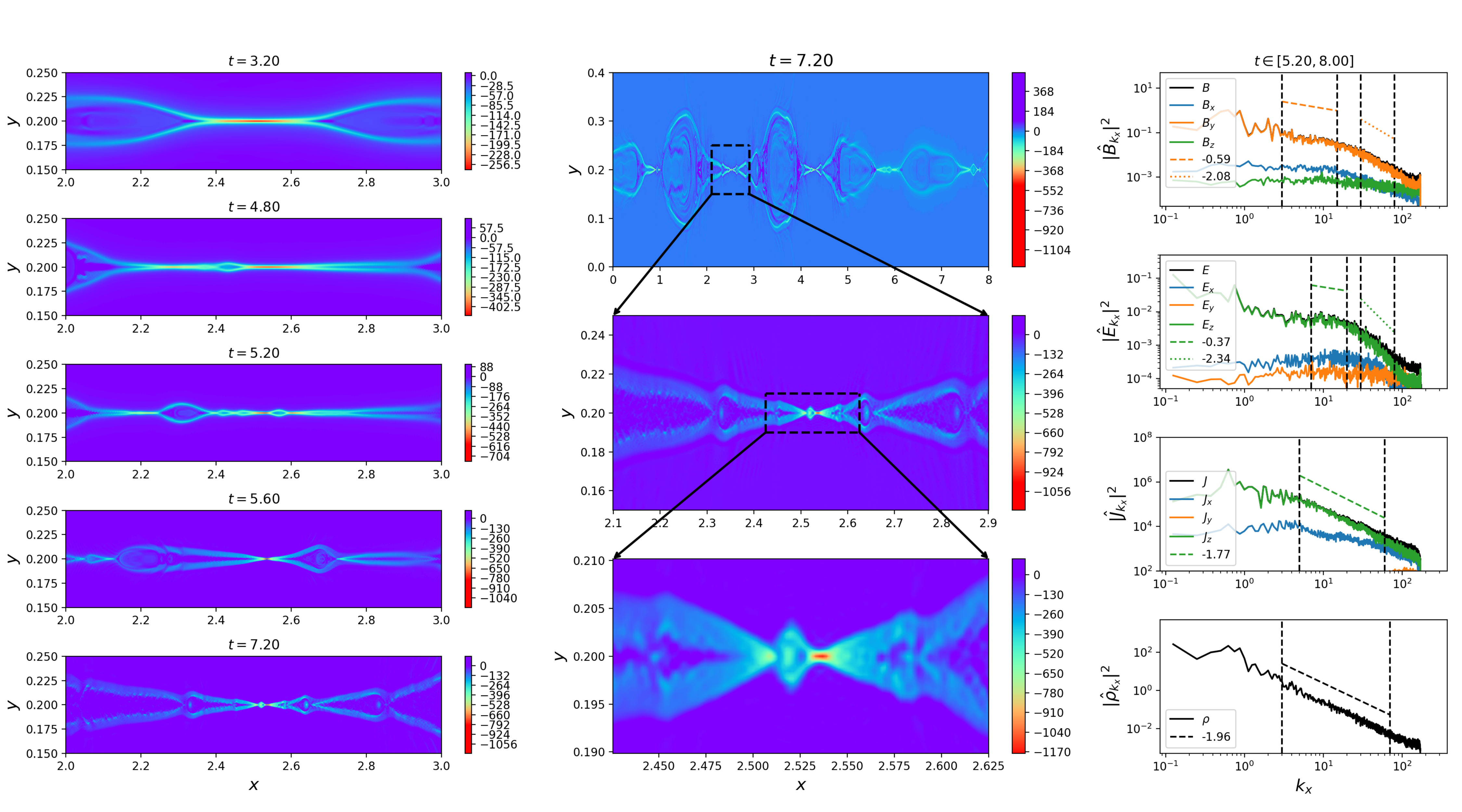}
\caption{Out-of-plane current density $J_z$ and the power spectra in Run 1. Left column: time evolution of $J_z$ inside the domain $x \in[2.0,3.0] $ and $y \in [0.15,0.25]$. Middle column: the hierarchic structure of the current sheet (color is $J_z$) at $t=7.20$. Top panel is the whole simulation domain. Middle and bottom panels are blow-ups as marked by the boxes and arrows. Right column: time-averaged power spectra of the magnetic field $\mathbf{B}$, the electric field $\mathbf{E}$, the current density $\mathbf{J}$ and the density $\rho$ calculated at the central line of the current sheet $y=0.2$ in log-log scale. The blue, orange and green lines represent the $x$, $y$ and $z$ components of the vector fields respectively and the black lines are the sum of the 3 components in the first three panels. The dashed and dotted lines are linear fits for the spectra of $B_y$, $E_z$, $J_z$ and $\rho$. The $k_x$ ranges for the fits are marked by the vertical dashed lines and the slopes of the fits are written in the legends. The time period for averaging the spectra is written at the top of the column. \label{fig:time_series_recur_structure_spectra_1}}
\end{figure}

\begin{figure}[ht!]
\centering
\includegraphics[scale=0.35]{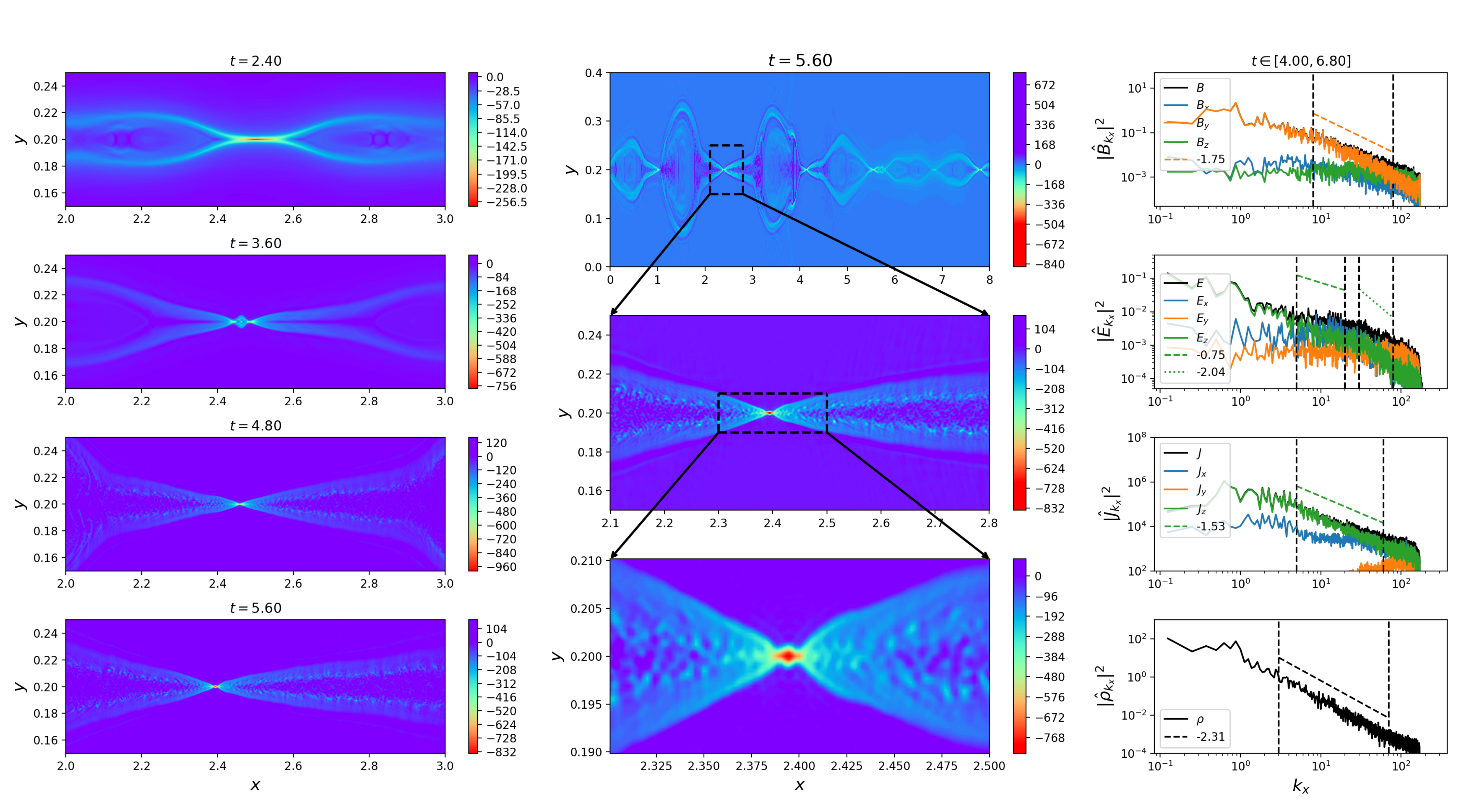}
\caption{Out-of-plane current density $J_z$ and the power spectra in Run 2. Left column: time evolution of $J_z$ inside the domain $x \in[2.0,3.0] $ and $y \in [0.15,0.25]$. Middle column: the hierarchic structure of the current sheet (color is $J_z$) at $t=5.60$. Top panel is the whole simulation domain. Middle and bottom panels are blow-ups as marked by the boxes and arrows. Right column: time-averaged power spectra of the magnetic field $\mathbf{B}$, the electric field $\mathbf{E}$, the current density $\mathbf{J}$ and the density $\rho$ calculated at the central line of the current sheet $y=0.2$ in log-log scale. The blue, orange and green lines represent the $x$, $y$ and $z$ components of the vector fields respectively and the black lines are the sum of the 3 components in the first three panels. The dashed and dotted lines are linear fits for the spectra of $B_y$, $E_z$, $J_z$ and $\rho$. The $k_x$ ranges for the fits are marked by the vertical dashed lines and the slopes of the fits are written in the legends. The time period for averaging the spectra is written at the top of the column. \label{fig:time_series_recur_structure_spectra_2}}
\end{figure}

\begin{figure}[ht!]
\centering
\includegraphics[scale=0.35]{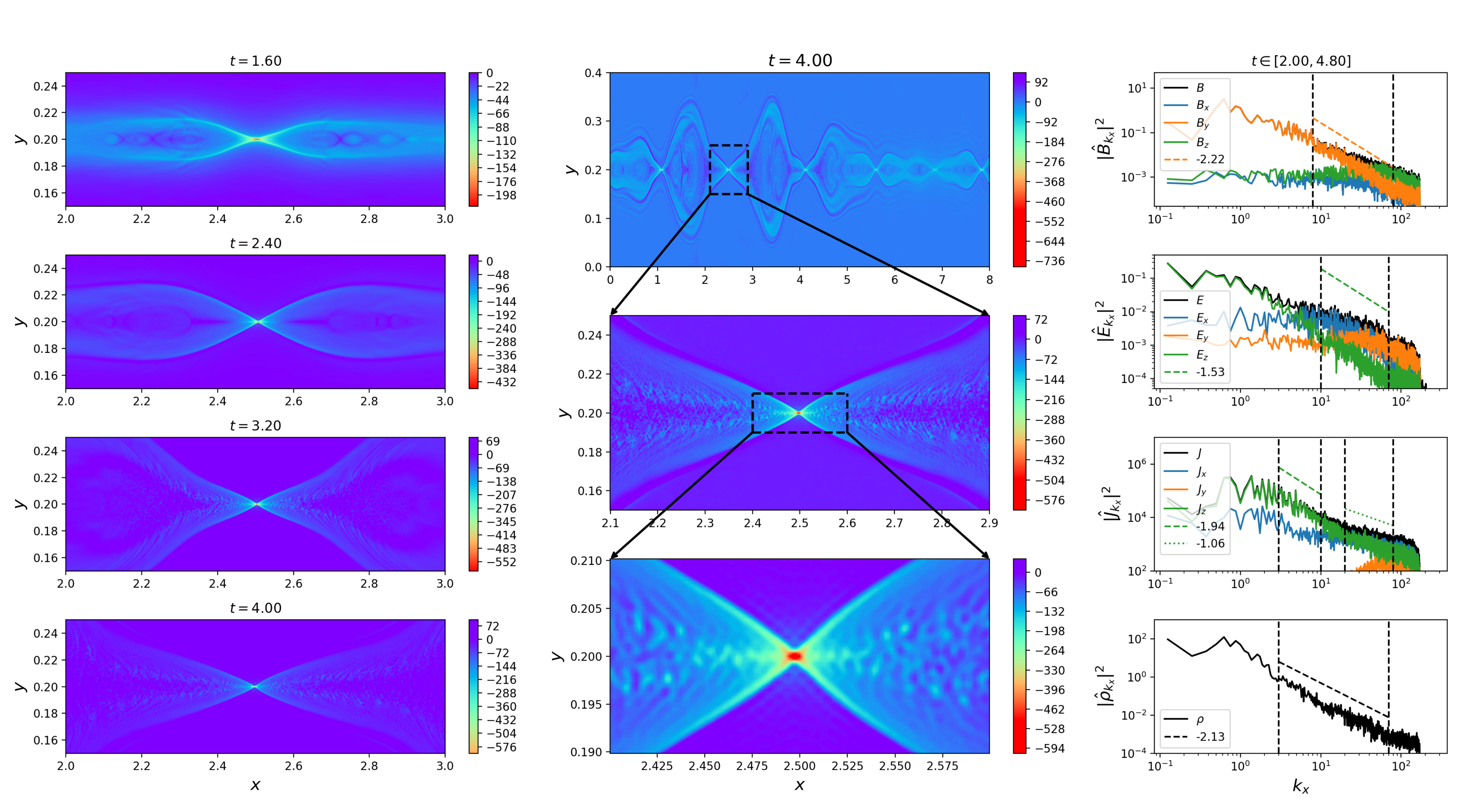}
\caption{Out-of-plane current density $J_z$ and the power spectra in Run 5. Left column: time evolution of $J_z$ inside the domain $x \in[2.0,3.0] $ and $y \in [0.15,0.25]$. Middle column: the hierarchic structure of the current sheet (color is $J_z$) at $t=4.00$. Top panel is the whole simulation domain. Middle and bottom panels are blow-ups as marked by the boxes and arrows. Right column: time-averaged power spectra of the magnetic field $\mathbf{B}$, the electric field $\mathbf{E}$, the current density $\mathbf{J}$ and the density $\rho$ calculated at the central line of the current sheet $y=0.2$ in log-log scale. The blue, orange and green lines represent the $x$, $y$ and $z$ components of the vector fields respectively and the black lines are the sum of the 3 components in the first three panels. The dashed and dotted lines are linear fits for the spectra of $B_y$, $E_z$, $J_z$ and $\rho$. The $k_x$ ranges for the fits are marked by the vertical dashed lines and the slopes of the fits are written in the legends. The time period for averaging the spectra is written at the top of the column. \label{fig:time_series_recur_structure_spectra_5}}
\end{figure}

Figure \ref{fig:time_series_recur_structure_spectra_1} is the similar plot with Figure \ref{fig:time_series_recur_structure_spectra_0} for Run 1, i.e. $d_i=0.001$. The evolution is similar with Run 0 at $t \leq 5$: a secondary current sheet is formed and stretched along $x$ direction with thickness $a_1  \approx 0.0016$. Then the secondary current sheet breaks into small plasmoids at $t\approx 5$, earlier than Run 0 as expected: tearing instability has larger growth rate with Hall term \cite{pucci2017}. Very soon after the break up of the secondary current sheet, a single X-point configuration, which is a typical structure of Hall-reconnection, is formed (between $t=5.2$ and $t=5.6$). But instead of a steady X-point structure, plasmoids are consecutively generated at the X-point. The bottom-left panel and the middle column of Figure \ref{fig:time_series_recur_structure_spectra_1} clearly show a chain of wedges on each side of the X-point at $t=7.2$ and these wedges are the ejected plasmoids. As shown in the bottom panel of the middle column, there is one plasmoid being generated and ejected toward the left at $x\approx 2.52$. Figure \ref{fig:time_series_recur_structure_spectra_2} and Figure \ref{fig:time_series_recur_structure_spectra_5} are  plots similar to those shown in Figure \ref{fig:time_series_recur_structure_spectra_0} but for Run 2 and Run 5, respectively. As $d_i$ increases, the Hall-reconnection stage is triggered faster and the plasmoid generation is suppressed. In Run 2 we can still see generation of plasmoids (see the second panel on the left column and the bottom panel on the middle column of Figure \ref{fig:time_series_recur_structure_spectra_2}) while for Run 5 there is no plasmoid observed. 

\begin{figure}[ht!]
\centering
\includegraphics[scale=0.8]{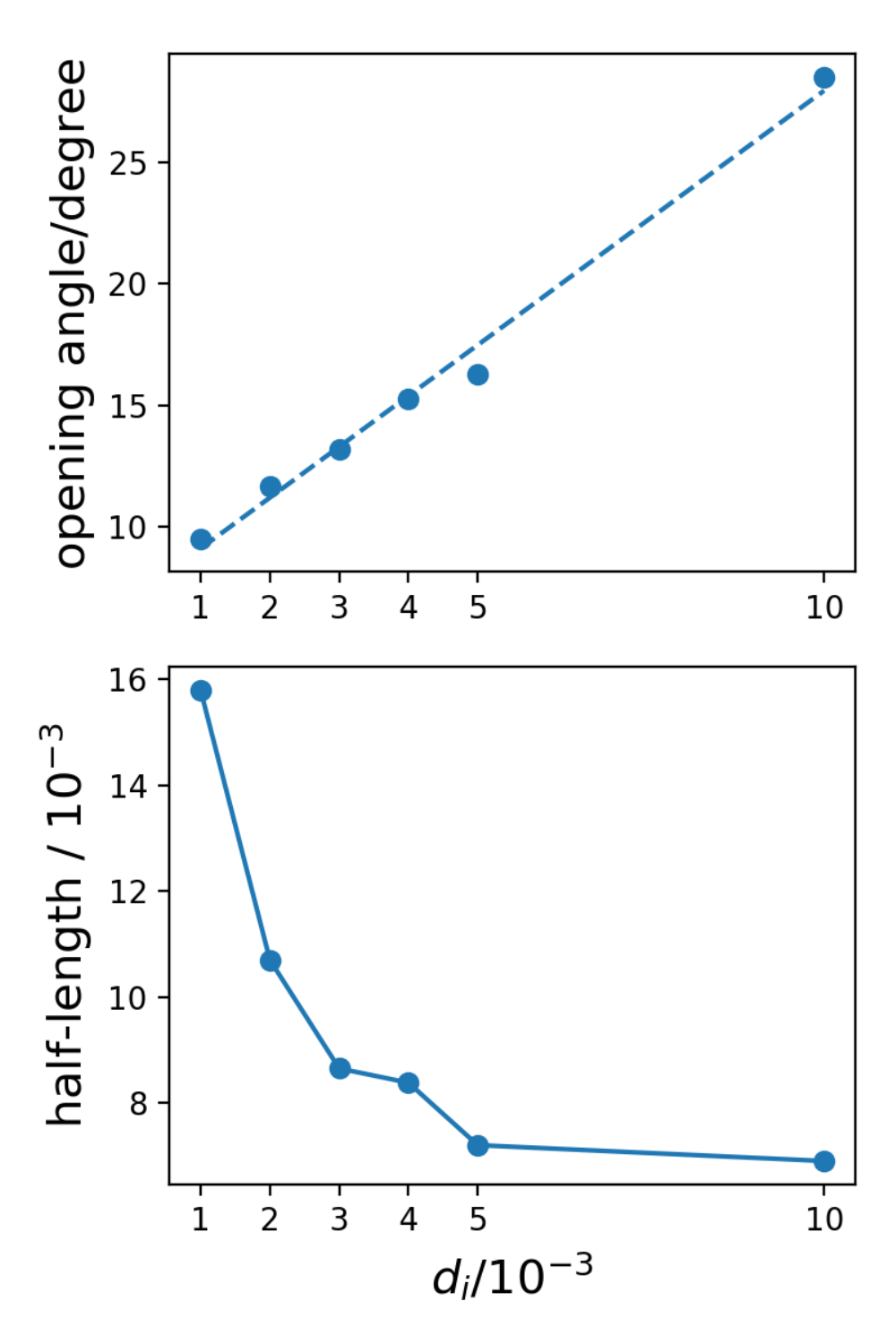}
\caption{Top: opening angle of the exhaust as a function of $d_i$. Bottom: half-length of the current sheet at the X-point as a function of $d_i$. \label{fig:open_angle_half_length}}
\end{figure}

By comparing the middle-bottom panels of Figure \ref{fig:time_series_recur_structure_spectra_1}-\ref{fig:time_series_recur_structure_spectra_5}, which show the simulation domains of the same size ($0.2 \times 0.02$), we observe that the opening angle of the exhaust increases and the size of the X-point shrinks as $d_i$ enlarges. In Figure \ref{fig:open_angle_half_length}, we show the opening angle of the exhaust and the half length of the current sheet at the X-point as functions of $d_i$. The two quantities are calculated for each of Run 1-5 and for Run 10, when the X-point structure is steady. The opening angle increases almost linearly with $d_i$ from about 10 degrees for $d_i=0.001$ to about 28 degrees for $d_i = 0.01$. The half-length of the current sheet decreases with $d_i$ quite fast for $d_i \leq 0.003$ and then begins to converge, possibly because the resolution $\Delta x \approx 0.002$ is approached.

\subsection{Reconnection rate and dissipation}
\begin{figure}[ht!]
\centering
\includegraphics[scale=0.6]{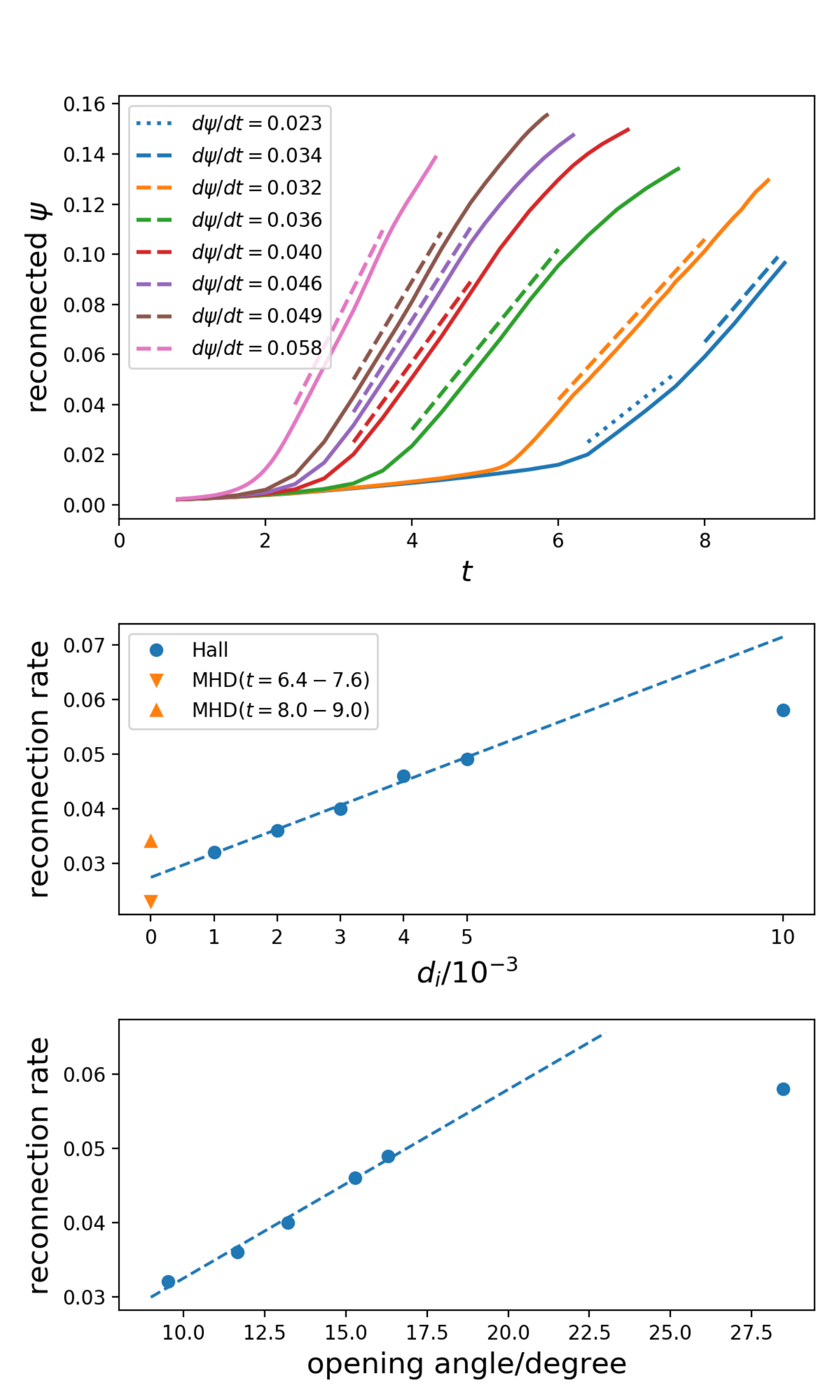}
\caption{Top: reconneced magnetic flux at the X-point as a function of time. Dashed lines are linear fits whose slopes are shown in the legend. Blue, orange, green, red, purple, brown and pink lines (from right to left) are Run 0, 1, 2, 3, 4, 5, 10 respectively. Middle: reconnection rate calculated by the linear fits as a function of $d_i$. Note that for Run 0 there are two growth rates corresponding to the two fast-growing stages as shown in the top panel. Bottom: reconnection rate as a function of the opening angle for Run 1-5 and 10. \label{fig:reconnected_flux_reconnection_rate}}
\end{figure}

In Figure \ref{fig:reconnected_flux_reconnection_rate} we plot in the top panel the reconnected magnetic flux $\psi$ as a function of time for all runs. The curves from right to left correspond to Run 0,1,2,3,4,5 and 10 respectively. For each run, we confine the calculation to be within the area $x \in[2.0,3.0] $ and $y \in [0.15,0.25]$. As the code evolves the $z$-component of the vector potential $\psi$ instead of the in-plane magnetic field, the reconnected magnetic flux is easily acquired by recording the value of $\psi$ at the X-point, defined as the point at which $\psi$ peaks. The time at which we begin to record $\psi$ is $t \lesssim 0.8$ after the dominant X-point in this area is formed. All runs go through a slowly-growing initial phase with slightly different reconnection rate which increases with $d_i$ and the initial phase corresponds to the evolution of the first tearing. For Run 1-5 and 10, fast reconnection is triggered simultaneously with the formation of the Hall-like configuration, e.g. at $t \sim 5$ for Run 1 and $t \sim 4$ for Run 2. The larger $d_i$, the earlier the fast-reconnection phase is entered. Then the reconnection rate remains steady during the fast-reconnecting stage. However, for Run 0, i.e. the MHD case, the process is different from the other runs. First, the reconnection rate increases when the secondary tearing happens, i.e. when the secondary current sheet breaks up at $t \sim 6$, instead of the formation of a Hall-like structure. Second, the reconnection does not maintain a steady rate throughout the fast-reconnection stage. At $t \ 8$, the reconnection rate increases once again due to the third or higher-order tearing (see bottom panel on the middle column of Figure \ref{fig:time_series_recur_structure_spectra_0}). The dashed lines in the top panel of Figure \ref{fig:reconnected_flux_reconnection_rate} are linear fits of the $\psi(t)$ curves and the slopes of the fitted lines are written in the legend. In the middle panel we plot the reconnection rate calculated from the linear fits as a function of $d_i$. For Run 1-5, the reconnection rate increases with $d_i$ roughly linearly and its value is around $0.03-0.05$. For Run 10, the reconnection rate is $0.058$, lower than the linear extrapolation of Run 1-5 as shown by the dashed line, indicating a convergence of the reconnection rate as $d_i$ increases. The reconnection rates of these runs are consistent with previous Hall-reconnection simulations \cite{ma2001}. The triangles correspond to the two reconnection rates before and after the second acceleration of the reconnection in Run 0. Note that the fastest reconnection rate in the MHD case is even larger than that of the $d_i=0.001$ run. If the resolution of the simulation is good enough to resolve smaller-scale current sheets, the reconnection rate is supposed to increase further. In the bottom panel of Figure \ref{fig:reconnected_flux_reconnection_rate} we plot the reconnection rate as a function of the opening angle for Run 1-5 and 10. This plot indicates that as the opening angle of the exhaust increases, the reconnection rate goes up when the opening angle is small but does not go up all the way: there is a maximum reconnection rate allowed in the system. \cite{liu2017} discussed this problem by a simple model where the micro-scale diffusion region is embedded inside the large-scale MHD region. As the exhaust opens up, the magnetic field right upstream the diffusion region is reduced because of the balance between the magnetic pressure gradient force and the magnetic tension force in the MHD-scale upstream region. As a consequence, the outflow speed and the reconnection rate are also reduced. In their estimate, the opening angle corresponding to the maximum reconnection rate is $\alpha = 2 \times \arctan{(0.31)} = 0.60 = 34^o$ which is not reached by our simulations but from the bottom panel of Figure \ref{fig:reconnected_flux_reconnection_rate} we can see the trend of a saturated reconnection rate as the opening angle increases toward $30^o$.

\begin{figure}[ht!]
\centering
\includegraphics[scale=0.5]{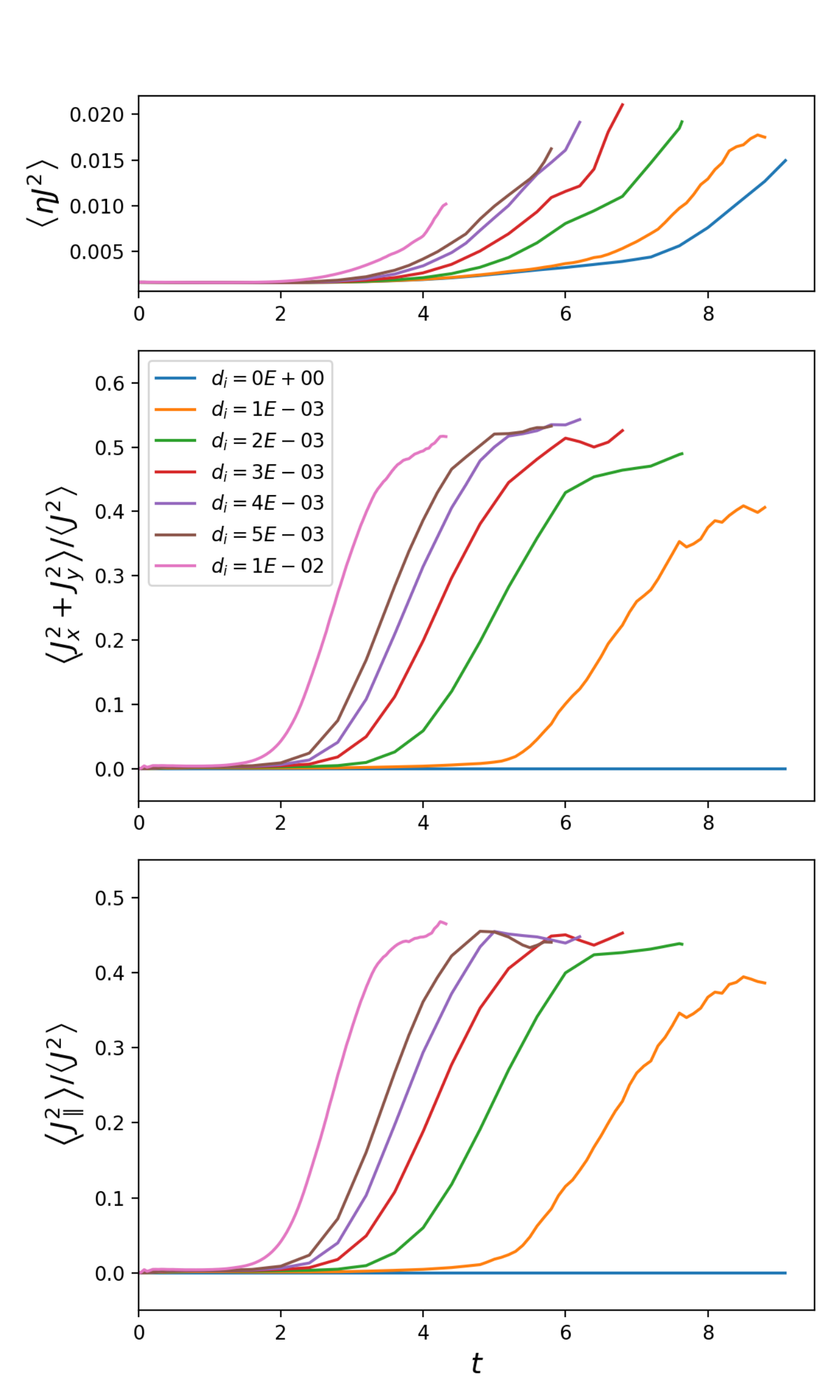}
\caption{Top: dissipation rate $\eta J^2$ averaged over the whole simulation domain. Middle: the ratio between the Hall-current dissipation rate $\eta \left( J_x^2 + J_y^2 \right)$ and the total dissipation rate $\eta J^2$. Bottom: the ratio between the parallel dissipation rate and the total dissipation rate. \label{fig:dissipation}}
\end{figure}

In the top panel of Figure \ref{fig:dissipation} we show the time evolution of the dissipation rate averaged over the whole simulation domain:
\begin{equation}
    D = \left<\eta J^2 \right> = \frac{1}{n_x n_y} \int \eta J^2  =\frac{1}{n_x n_y} \int \mathbf{J} \cdot \mathbf{E^\prime}
\end{equation}
where 
\begin{equation}
    \mathbf{E^\prime} = \mathbf{E} + \mathbf{u_e} \times \mathbf{B}
\end{equation}
is the electric field in electron frame. The evolution of the dissipation rate is different from that of the reconnected magnetic flux (top panel of Figure \ref{fig:reconnected_flux_reconnection_rate}). The reconnection rate increases abruptly when either the second tearing or the Hall-reconnection is triggered while the dissipation rate increases more smoothly with time due to the averaging over the whole domain. In addition, the dissipation rate at the end of the simulation decreases with $d_i$ for $d_i \ge 0.003$: as $d_i$ increases, the total time of reconnection shortens due to the finite magnetic flux in the simulation domain and faster reconnection rate, leading to the decrease of the amplitude of the current density at the end of the simulation. In the middle panel we show the ratio between the Hall-current dissipation rate
\begin{equation}
    D_{xy} = \left< \eta \left( J_x^2 + J_y^2 \right) \right>
\end{equation}
and the total dissipation rate $D$ as a function of time for each run. For the MHD case this ratio remains 0 as there is no Hall current while for $d_i \neq 0$, the significance of the Hall-current dissipation grows with time once the Hall-reconnection stage initiates and the ratio tends to saturate. Even for $d_i=0.001$, the Hall-current dissipation accounts for $\sim 40 \%$ of the total dissipation at the end of the simulation. As $d_i$ increases, this ratio converges to $\sim 55 \%$, i.e. the dissipation by the Hall current becomes more important than that of the out-of-plane current. We also check the ratio between the parallel dissipation rate $\eta J_\parallel^2$ and the total dissipation rate where $J_\parallel = \mathbf{J} \cdot \mathbf{B} / B$ and show this ratio in the bottom panel. It evolves similarly with the Hall-dissipation ratio and converges to a value $\sim 45\%$ as $d_i$ increases.

\subsection{Power spectra}\label{sec:power_spectra}
It is expected that the nonlinear recursive generation of plasmoids leads to an energy cascade from large to small scales, leading to well-developed power-law spectra. The development of turbulence with
an extended inertial range is important both to energization \cite{lu2019} and scattering of particles \cite{giacalone2000}. In the MHD case, a phenomenological model based on the recursive tearing mode developed by \cite{tenerani2019} leads to a spectrum $\left|B_{y}(k_x)\right|^2 \propto k_x^{-4/5}$. In the kinetic regime, i.e. below the ion inertial length, the power spectra of the turbulence generated are different due to the presence of kinetic mechanisms, as discussed below. We inspect the power spectra of various quantities: the magnetic field $\mathbf{B}$, the electric field $\mathbf{E}$, the current density $\mathbf{J}$ and the density $\rho$. The right columns of Figure \ref{fig:time_series_recur_structure_spectra_0}-\ref{fig:time_series_recur_structure_spectra_5} are the spectra for Run 0, 1, 2, 5 respectively. The spectra are calculated at the central line of the current sheet $y=0.2$ and are averaged over the time periods of steady and fast reconnection. The time periods for averaging the spectra are written at the top of the columns. The blue, orange and green lines are the $x$, $y$ and $z$ components of the vector fields respectively. The black lines in the top three panels are the sum of the 3 components. Note that in some panels the total power spectrum is covered by the dominant component, e.g. in Run 0 (Figure \ref{fig:time_series_recur_structure_spectra_0}) $E_z$ overlap with the total power of $\mathbf{E}$ as $E_x$ and $E_y$ are exactly 0 in this run. We fit the powers of $B_y$, $E_z$, $J_z$ and $\rho$ over $k$ ranges marked by the vertical dashed lines where obvious exponential relations exist. The fits are shown by the dashed and dotted lines whose slopes are written in the legends.

\begin{figure}[ht!]
\centering
\includegraphics[scale=0.6]{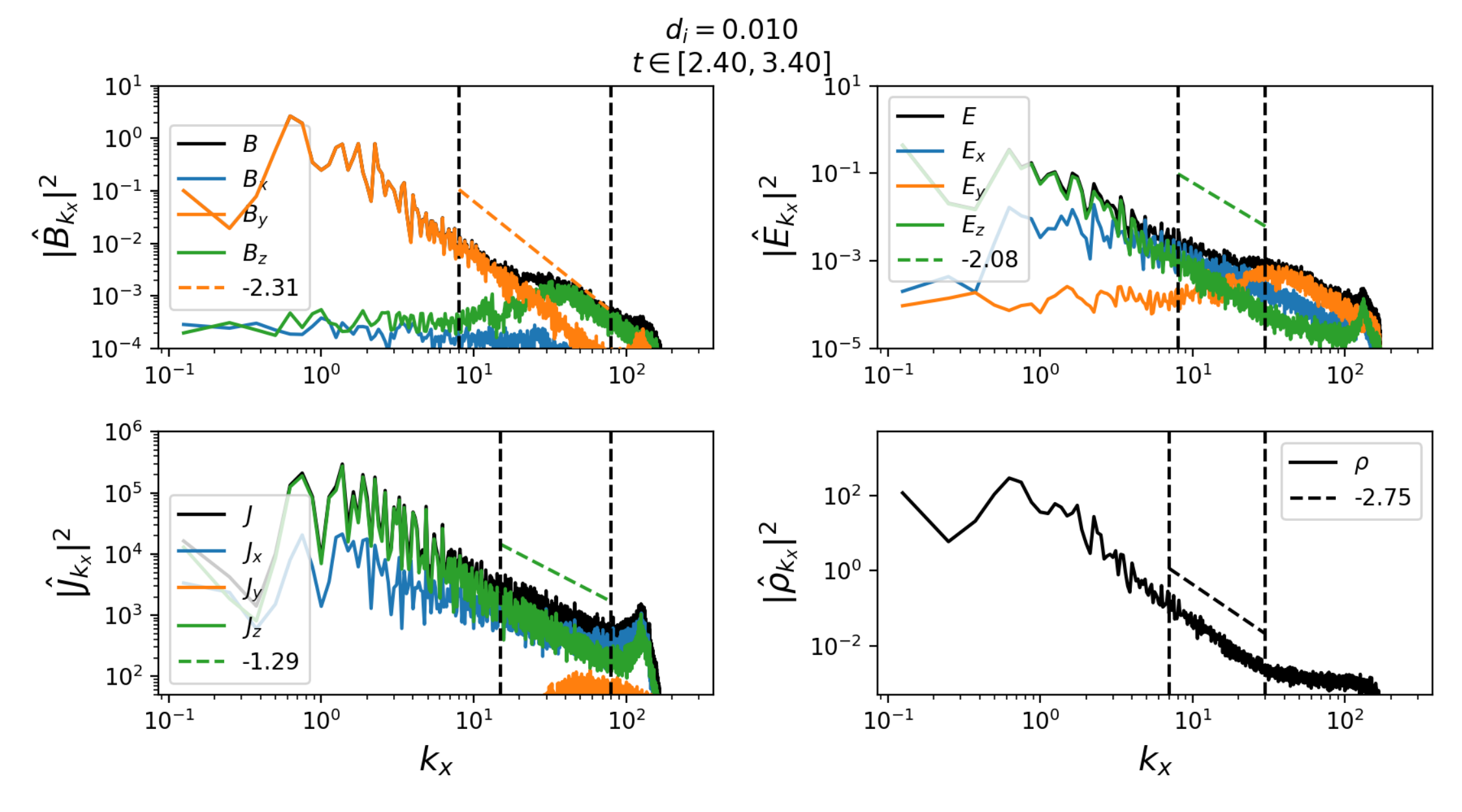}
\caption{Power spectra averaged from $t=2.4$ to $t=3.4$ for Run 10 in log-log scale. The spectra are calculated at the central line of the current sheet $y=0.2$. From top-left to bottom-right panels are the magnetic field $\mathbf{B}$, the electric field $\mathbf{E}$, the current density $\mathbf{J}$ and the density $\rho$. The blue, orange and green lines are the $x$, $y$ and $z$ components of the vector fields respectively. The black lines are the sum of the 3 components in panels for $\mathbf{B}$, $\mathbf{E}$ and $\mathbf{J}$. The dashed lines are the fits for $B_y$, $E_z$, $J_z$ and $\rho$ over $k_x$ ranges marked by the vertical dashed lines. The slopes of the fits are shown in the legends. \label{fig:power_spectra_10}}
\end{figure}

We first look at the power spectra of $\mathbf{B}$ in Figure \ref{fig:time_series_recur_structure_spectra_0}-\ref{fig:time_series_recur_structure_spectra_5}. At the center of the current sheet, the dominant component is $B_y$ as shown by the orange lines. We see that the power spectra peak around $k_x \approx 0.6-0.8$ in all of the runs shown in Figure \ref{fig:time_series_recur_structure_spectra_0}-\ref{fig:time_series_recur_structure_spectra_5}. This peak corresponds to the 5-6 large-scale plasmoids in all these runs as shown in the top panels on the middle columns of Figure \ref{fig:time_series_recur_structure_spectra_0}-\ref{fig:time_series_recur_structure_spectra_5}. Note that the number of the large plasmoids is a nonlinear effect related probably to the initial condition and the domain size rather than the fastest-growing linear mode. For Run 0 and Run 1, a clear two-segment power spectrum of $B_y$ is seen with the break point at $k_x \approx 20$, i.e. $\lambda \approx 0.05$. This value of $\lambda$ is close to the length of the third-order current sheets in Run 0 (see the panel $t=6.40$ on the left column of Figure \ref{fig:time_series_recur_structure_spectra_0}) and the length of the current sheet at the X-point in Run 1, which is around $0.03-0.04$ (see the bottom panel of Figure \ref{fig:open_angle_half_length}). Below the break point, the spectral index (defined as the absolute value of the slope) is smaller than 1, while above the break point the spectral index is very close to 2. Especially, in Run 0 the slope of $B_y$ spectrum below the break point is approximately $-4/5$, the value estimated by \cite{tenerani2019}, indicating that in the low-$k_x$ range the spectra are controlled by the recursive plasmoid generation. In high-$k_x$ range the $-2$ slope is mainly due to the discontinuities near the boundaries of the plasmoids. For Run 2, the $B_y$ spectrum becomes smooth. Although we can still see that the slope of the spectrum changes with $k_x$, we are unable to find a break point. For Run 5, the $B_y$ spectrum is almost a straight line with a slope $-2.22$. The change of the shape of $B_y$ spectra is related to the formation of the X-point structure due to the Hall effect. As $d_i$ increases, the recursive growth of plasmoids is suppressed. As a result, the shallow part of the $B_y$ spectrum, which represents the recursive generation of plasmoids, is gradually taken over by the steeper spectrum indicating more discontinuities/shocks produced in the Hall cases. The second panels on right columns of Figure \ref{fig:time_series_recur_structure_spectra_0}-\ref{fig:time_series_recur_structure_spectra_5} show the power spectra of $\mathbf{E}$ and we calculate the spectral indices for $E_z$, the out-of-plane electric field. For Run 0 and Run 1, $E_z$ is dominant and the power spectrum of $E_z$ consists of two segments like $B_y$ though the high-k part of $E_z$ spectrum is steeper than $B_y$ with slopes around  -2.3. For Run 2, $E_z$ still shows a two-segment power spectrum while $B_y$ does not. For Run 5, $E_z$ power spectrum is almost a straight line with slope $-1.53$, flatter than that of $B_y$. We also notice that, as $d_i$ increases, the energy in the Hall electric field $E_x$ (the blue lines) exceeds the energy in $E_z$ at small scales (see Figure \ref{fig:time_series_recur_structure_spectra_2} and \ref{fig:time_series_recur_structure_spectra_5}). In Run 5, $E_x$ dominates the electric energy at scales $k_x > 5$, i.e. the Hall electric field is the major contributor to the electric energy at scales $\lambda < 0.2$. The third panels of Figure \ref{fig:time_series_recur_structure_spectra_0}-\ref{fig:time_series_recur_structure_spectra_5} show the power spectra of the current density $\mathbf{J}$ and we calculate the spectral indices for $J_z$. We see that $J_z$ dominates the power of $\mathbf{J}$ in all runs except at quite small scales in Run 2 and Run 5. For Run 0-2, the $J_z$ power spectra are similar with spectral indices around $1.5-1.8$ while for Run 5, the $J_z$ power spectrum has a break point at $k_x \sim 10-20$ below which the spectral index is around 2 and above which the spectral index is around 1. The bottom panels of Figure \ref{fig:time_series_recur_structure_spectra_0}-\ref{fig:time_series_recur_structure_spectra_5} show the power spectra of the density $\rho$ and we see that the 4 runs are very similar with slopes around $-2$, indicating that the Hall effect does not change the compressional properties significantly in this $d_i$ range. For comparison, we plot the spectra for Run 10 in Figure \ref{fig:power_spectra_10}. We see that $\mathbf{B}$, $\mathbf{E}$ and $\mathbf{J}$ show similar power spectra with Run 5 though $E_z$ spectrum is steeper and it is hard to calculate the spectral index for $J_z$ at large scales due to the strong oscillation. The density has a significantly steeper power spectrum in Run 10 with slope $-2.75$, indicating a stronger damping of the compressible fluctuations in this run.

\section{Summary}\label{sec:summary}
In this work, we have carried out a series of 2.5D Hall-MHD simulations to study the transition from resistive-MHD reconnection to Hall-reconnection. We show that as the ion inertial length increases from 0 to above the inner-layer thickness of the ideal tearing mode, the evolution and the structure of the current sheet change significantly. For $d_i=0$ (or $d_i \ll \delta_{TI} $ where $\delta_{TI}$ is the inner layer thickness of the tearing mode), the reconnection is purely plasmoid-dominant, consistent with the recursive reconnection picture \cite[e.g.]{shibata2001,tenerani2015}. When $d_i \gg \delta_{TI}$, the reconnection evolves into the Hall-dominant regime soon after the first tearing: a single X-point with exhaust opened up. When $d_i \lesssim \delta_{TI}$, we see an intermediate state: an X-point with finite length is formed and plasmoids continue to be generated and ejected out from the X-point. The steady-nonlinear reconnection rate for Hall reconnection is around $0.03-0.06$, increasing with $d_i$ and  with a trend toward a maximum value around $0.06$ in our simulations. The resistive-MHD case is also able to give fast nonlinear reconnection rate ($0.034$) and this rate increases with time as higher-order tearing happens. The Hall effect changes the energy dissipation significantly. The Hall current system accounts for a large portion ($40 \%$) of the total dissipation even in our Run 1 where the reconnection is the intermediate plasmoid+Hall state and as $d_i$ increases this portion converges to a value $\gtrsim 50\%$. Last, the Hall term changes the power spectra of various quantities a lot, especially the electromagnetic fields $\mathbf{B}$ and $\mathbf{E}$. For instance, the reconnecting magnetic field component $B_y$ shows two-segment power spectra with slopes around $-0.6 \sim -0.8$ at small $k_x$ and slopes around $-2$ at large $k_x$ in Run 0 and 1 while in Run 2-5 and 10 its power spectra are one-segment and become steeper and steeper as $d_i$ increases. This change of the spectra is related to the suppression of the recursive plasmoid generation and the formation of Petschek-like X-points as the Hall effect becomes stronger.

In the Hall-MHD model, the temperature anisotropy is absent but it is known that it develops in ion scale current sheets \cite{drake2009}. This temperature anisotropy may facilitate the formation of slow shocks so that the structure of the reconnection becomes Petschek-like more easily \cite{liu2011}. Besides, the formed temperature anisotropy may lead to the growth of various instabilities, e.g. fire-hose instability capable of perturbing the outflow regions \cite{hietala2015}.

\cite{cassak2010} propose that ``bistability'' exists in Hall-MHD reconnection: the Sweet-Parker and Hall solutions are both stable. However, their result is not applicable to many astrophysical environments where the resistivity is so small that the Sweet-Parker current sheet cannot form in the first place. \cite{huang2011} observe the bounce between the Hall-like X-point and the Sweet-Parker-like long current sheet in their simulations of two coalescing magnetic islands. However, in our simulations we do not observe such phenomenon. In our Run 1 we see a stable status where plasmoids are generated out of the X-point continuously but we do not see the transition from the X-point structure back to a long current sheet. The reason may be that our initial configuration is a pressure-balanced Harris current sheet rather than two coalescing magnetic islands. In the simulations by \cite{huang2011}, the large coalescing islands may possibly ``squeeze'' the current sheet in between and destroy the opened-up exhaust of the Hall-like structure. 

In the solar wind and the solar corona, the relation between $d_i$, $a_0$ and $\delta_{TI}$ may be similar to the setup in this work \cite[see]{pucci2017}. Thus our study may help us understand the reconnection process in these environments. As the Parker Solar Probe (PSP) is collecting data in the inner heliosphere and solar corona , it will be of great interest to combine the numerical study with the PSP data analysis in the future. In addition, the simulations in the current study are 2.5D and it is necessary to extend this work to 3D.

\section{Acknowledgments}
We would like to thank Fulvia Pucci and Yi Qi for many useful discussions. Chen Shi would also like to thank Kun Zhang for her support. We acknowledge support from the NASA LWS Grant No. NNX15AF34G and
the NSF-DOE Partnership in Basic Plasma Science and
Engineering Award No. 1619611. This work used the Extreme Science and Engineering Discovery Environment (XSEDE) \cite{towns2014} Comet at the San Diego Supercomputer Center through allocation TG-AST160007. XSEDE is supported by National Science Foundation grant number ACI-1548562.

\bibliographystyle{unsrt}  


\end{document}